\documentstyle[aps,epsfig]{revtex}

\begin{document}
\title{Transversity and intrinsic motion of the constituents}
\author{A.V. Efremov$^{1}$, O.V. Teryaev$^{1}$ and P. Z\'{a}vada$^{2}$}
\address{$^{1}$Bogoliubov Laboratory of Theoretical Physics, JINR, 141980\\
Dubna, Russia\\
$^{2}$Institute of Physics, Academy of Sciences of the Czech Republic, \\
Na Slovance 2, CZ-182 21 Prague 8}
\date{August 17, 2004}
\maketitle

\begin{abstract}
The probabilistic model of parton distributions, previously developed by one
of the authors, is generalized to include the transversity distribution.
When interference effects are attributed to quark level only, the intrinsic
quark motion produces the transversity, which is about twice as large as the
usual polarized distribution. The applicability of such a picture is
considered and possible corrections, accounting for interference effects at
the parton-hadron transition stage are discussed.
\end{abstract}

\section{Introduction}

Nucleon spin functions represent a sensitive tool for understanding the
nucleon internal structure in the language of QCD. Up to this day we have
accumulated a very good knowledge of the nucleon spin functions $g_{1}$\ and 
$g_{2}$, which were measured in deep inelastic scattering \cite{e142} -\cite%
{her2}. A further important and interesting quark spin distribution function
is the transversity, the third non-diagonal element of the quark spin
density matrix. Transversity is not accessible from the measuring of deep
inelastic scattering, since it corresponds to the helicity flip amplitude.
Its measuring is more complicated and that is the reason, why some more
accurate and complete experimental data on the transversity are still
missing. However, the recent and/or future data from the experiments HERMES
(DESY-Hamburg), CLAS (JLab), COMPASS (CERN-Geneva) and RHIC (Brookhaven
National Laboratory) could be interpreted also in terms of the transversity %
\cite{san2000fh} -\cite{efre2001cz}. For the present status of research in
both theory and experiment, see e.g. \cite{dis04} \ and overview \cite%
{barone}.

In Refs. \cite{zav4},\cite{zav5} the probabilistic, covariant quark-parton
model (QPM), in which intrinsic quark motion with spheric symmetry is
consistently taken into account, was developed by one of us (P.Z.). It was
shown that such a model nicely reproduces some well-known sum rules and
gives a very reasonable agreement with experimental data on the spin
structure functions $g_{1}$ and $g_{2}$. Assuming $SU(6)$ symmetry, a
calculation was done from the input on unpolarized valence quark
distributions $q_{V}$. The aim of this paper is to extend this model also
for description and calculation of the transversity distribution.

\section{Transversity}

\begin{figure}[tbp]
\begin{center}
\epsfig{file=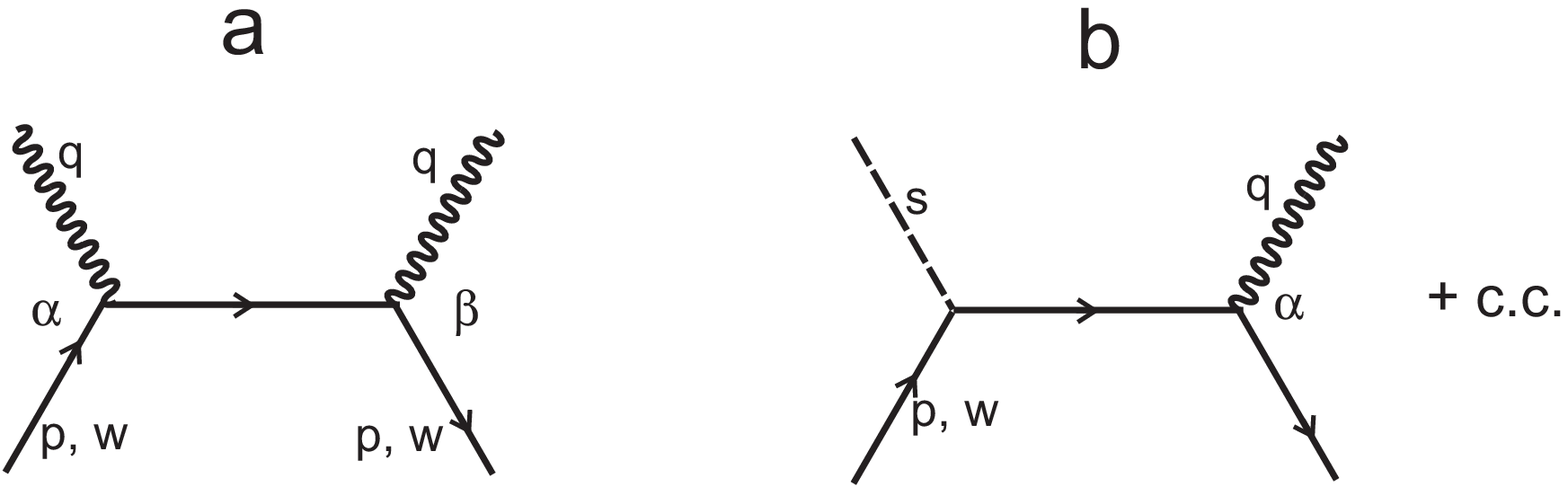, height=5cm}
\end{center}
\caption{Diagram related to deep-inelastic scattering (a) and the
transversity (b), see text.}
\label{fg1}
\end{figure}

First, let us shortly summarize, how the spin structure functions $%
g_{1},g_{2}$ were calculated in the paper \cite{zav4}. The antisymmetric
part of the tensor related to the photon absorption by a single quark reads:%
\begin{equation}
t_{\alpha \beta }=m\varepsilon _{\alpha \beta \lambda \sigma }q^{\lambda
}w^{\sigma },  \label{tt1}
\end{equation}%
where $q,m,w$ are the photon momentum, quark mass and polarization vector;
the corresponding handbag diagram is in Fig. \ref{fg1}a. Then it was shown
that the corresponding tensor related to the target (proton) consisting of
quasifree quarks can be written as%
\begin{equation}
T_{\alpha \beta }^{(A)}=\varepsilon _{\alpha \beta \lambda \sigma
}q^{\lambda }\frac{m}{2Pq}\int H\left( \frac{pP}{M}\right) w^{\sigma }\delta
\left( \frac{pq}{Pq}-x\right) \frac{d^{3}p}{p_{0}};\qquad x=\frac{Q^{2}}{2Pq}%
,  \label{cr17}
\end{equation}%
where $M$ is the proton mass, $p$ and $P$ are the quark and proton momenta.
The distribution $H$ is the difference of the quark distributions with
opposite spin projections. In the proton rest frame one can write%
\begin{equation}
H(p_{0})=G_{+}(p_{0})-G_{-}(p_{0}).  \label{tt2}
\end{equation}%
For the time being, if not stated otherwise, we consider the quark charge
equals unity. Further, we showed that the covariant form of the quark
polarization vector reads%
\begin{equation}
w^{\sigma }=AP^{\sigma }+BS^{\sigma }+Cp^{\sigma },  \label{cr15}
\end{equation}%
where $S$ is the proton polarization vector and%
\begin{equation}
A=-\frac{pS}{pP+mM},\qquad B=1,\qquad C=\frac{M}{m}A.  \label{cr16}
\end{equation}%
Finally, in the last step the functions $g_{1},g_{2}$ were extracted from
the tensor (\ref{cr17}). In the approximation 
\begin{equation}
Q^{2}\gg 4M^{2}x^{2}  \label{tt2a}
\end{equation}%
and identifying the beam direction with coordinate 1 in the proton rest
frame, we obtain

\begin{eqnarray}
g_{1}(x) &=&\frac{1}{2}\int H(p_{0})\left( m+p_{1}+\frac{p_{1}^{2}}{p_{0}+m}%
\right) \delta \left( \frac{p_{0}+p_{1}}{M}-x\right) \frac{d^{3}p}{p_{0}},
\label{tt3} \\
g_{2}(x) &=&-\frac{1}{2}\int H(p_{0})\left( p_{1}+\frac{p_{1}^{2}-p_{T}^{2}/2%
}{p_{0}+m}\right) \delta \left( \frac{p_{0}+p_{1}}{M}-x\right) \frac{d^{3}p}{%
p_{0}},  \label{tt4}
\end{eqnarray}%
which implies%
\begin{equation}
g_{T}(x)=g_{1}(x)+g_{2}(x)=\frac{1}{2}\int H(p_{0})\left( m+\frac{p_{T}^{2}/2%
}{p_{0}+m}\right) \delta \left( \frac{p_{0}+p_{1}}{M}-x\right) \frac{d^{3}p}{%
p_{0}}.  \label{ta14}
\end{equation}%
Since%
\begin{eqnarray*}
m+p_{1}+\frac{p_{1}^{2}}{p_{0}+m} &=&m-p_{0}+p_{0}+p_{1}+\frac{p_{1}^{2}}{%
p_{0}+m}=p_{0}+p_{1}-\frac{\left( p_{0}-m\right) \left( p_{0}+m\right)
-p_{1}^{2}}{p_{0}+m} \\
&=&p_{0}+p_{1}-\frac{{\bf p}^{2}-p_{1}^{2}}{p_{0}+m}=p_{0}+p_{1}-\frac{%
p_{T}^{2}}{p_{0}+m},
\end{eqnarray*}%
the $\delta -$ function allows a useful alternative representation of $g_{1} 
$:

\begin{equation}
g_{1}(x)=\frac{1}{2}\int H(p_{0})\left( Mx-\frac{p_{T}^{2}}{p_{0}+m}\right)
\delta \left( \frac{p_{0}+p_{1}}{M}-x\right) \frac{d^{3}p}{p_{0}}.
\label{tt4a}
\end{equation}%
Let us also remark that in the paper \cite{zav1} we showed that a similar
approach for the corresponding symmetric part of the proton tensor gives the
unpolarized structure function%
\begin{equation}
f_{1}(x)=\frac{F_{2}(x)}{x}=Mx\int \left( G_{+}(p_{0})+G_{-}(p_{0})\right)
\delta \left( \frac{p_{0}+p_{1}}{M}-x\right) \frac{d^{3}p}{p_{0}}.
\label{tt5}
\end{equation}

Now, if one assumes the same spheric shape of the distributions $G_{\pm }$
for both opposite polarizations, then the corresponding probabilities can be
parameterized as%
\begin{equation}
G_{+}=G(p_{0})\cos ^{2}\left( \eta /2\right) ,\qquad G_{-}=G(p_{0})\sin
^{2}\left( \eta /2\right) ,\qquad 0\leq \eta \leq \pi ,  \label{tt6}
\end{equation}%
so for $\eta =0(\pi )$ we have a pure state with the polarization $+(-)$.
For example, in the case of $SU(6)$ we have $\cos \eta =2/3(-1/3)$ for $u(d)$
quarks. The last relations imply%
\begin{equation}
G_{+}(p_{0})+G_{-}(p_{0})=G(p_{0}),\qquad
G_{+}(p_{0})-G_{-}(p_{0})=G(p_{0})\cos \eta ,  \label{tt7}
\end{equation}%
so the relations (\ref{tt4}) - (\ref{tt5}) can be rewritten as%
\begin{eqnarray}
g_{1}(x) &=&\frac{1}{2}\cos \eta \int G(p_{0})\left( Mx-\frac{p_{T}^{2}}{%
p_{0}+m}\right) \delta \left( \frac{p_{0}+p_{1}}{M}-x\right) \frac{d^{3}p}{%
p_{0}},  \label{tt8} \\
g_{2}(x) &=&-\frac{1}{2}\cos \eta \int G(p_{0})\left( p_{1}+\frac{%
p_{1}^{2}-p_{T}^{2}/2}{p_{0}+m}\right) \delta \left( \frac{p_{0}+p_{1}}{M}%
-x\right) \frac{d^{3}p}{p_{0}},  \label{tt9} \\
g_{T}(x) &=&\frac{1}{2}\cos \eta \int G(p_{0})\left( m+\frac{p_{T}^{2}/2}{%
p_{0}+m}\right) \delta \left( \frac{p_{0}+p_{1}}{M}-x\right) \frac{d^{3}p}{%
p_{0}},  \label{tt9a} \\
f_{1}(x) &=&Mx\int G(p_{0})\delta \left( \frac{p_{0}+p_{1}}{M}-x\right) 
\frac{d^{3}p}{p_{0}}.  \label{tt10}
\end{eqnarray}

Here, in the same approach, we shall try to calculate the transversity.
Generally, transversity may be related to the auxiliary polarized process
described by the interference of vector and scalar currents \cite{GJJ}, \cite%
{IK}, so that the respective quark tensor carries only one Lorentz index.
The simplest handbag diagram in Fig. \ref{fg1}b corresponds to the expression%
\begin{equation}
\tau _{\alpha }=\varepsilon _{\alpha \beta \lambda \sigma }p^{\beta
}q^{\lambda }w^{\sigma },  \label{tt11}
\end{equation}%
which will be used instead of the tensor (\ref{tt1}). In the next step we
integrate this vector equally as the tensor in (\ref{cr17}). Here we assume
for the time being, that due to rotational symmetry in the proton rest
frame, the transversity distribution is generated by the same function $H$
as that in the case of the longitudinal one: 
\begin{equation}
T_{\alpha }=\varepsilon _{\alpha \beta \lambda \sigma }q^{\lambda }\frac{1}{%
2Pq}\int H\left( \frac{pP}{M}\right) p^{\beta }w^{\sigma }\delta \left( 
\frac{pq}{Pq}-x\right) \frac{d^{3}p}{p_{0}}.  \label{n1}
\end{equation}%
Obviously only the terms $S^{\sigma },P^{\sigma }$ from the vector (\ref%
{cr15}) contribute here, i.e.,%
\begin{equation}
T_{\alpha }=\varepsilon _{\alpha \beta \lambda \sigma }q^{\lambda }\frac{1}{%
2Pq}\int H\left( \frac{pP}{M}\right) \left( S^{\sigma }-\frac{pS}{pP+mM}%
P^{\sigma }\right) p^{\beta }\delta \left( \frac{pq}{Pq}-x\right) \frac{%
d^{3}p}{p_{0}}.  \label{n2}
\end{equation}%
Now we take the proton rest frame and assume%
\begin{eqnarray}
S &=&(0,0,1,0),  \label{n3} \\
q &=&(\nu ,\left| {\bf q}\right| ,0,0),  \label{n4} \\
P &=&(M,0,0,0),  \label{na4}
\end{eqnarray}%
then \ one can check that only $T_{3}\neq 0$,%
\begin{equation}
T_{3}=\frac{1}{2M\nu }\int H\left( p_{0}\right) \left( p_{0}\left| {\bf q}%
\right| -p_{1}\nu -\left| {\bf q}\right| \frac{p_{2}^{2}}{p_{0}+m}\right)
\delta \left( \frac{p_{0}\nu -p_{1}\left| {\bf q}\right| }{M\nu }-x\right) 
\frac{d^{3}p}{p_{0}}.  \label{n5}
\end{equation}%
Further, if we again assume approximation (\ref{tt2a}), then 
\begin{equation}
\frac{\left| {\bf q}\right| }{\nu }=\sqrt{1+4M^{2}x^{2}/Q^{2}}\rightarrow 1
\label{n6}
\end{equation}%
and%
\begin{eqnarray}
T_{3} &=&\frac{1}{2}\int H\left( p_{0}\right) \left( \frac{p_{0}-p_{1}}{M}-%
\frac{p_{2}^{2}}{M\left( p_{0}+m\right) }\right) \delta \left( \frac{%
p_{0}-p_{1}}{M}-x\right) \frac{d^{3}p}{p_{0}}  \label{n7} \\
&=&\frac{1}{2}\int H\left( p_{0}\right) \left( x-\frac{p_{T}^{2}/2}{M\left(
p_{0}+m\right) }\right) \delta \left( \frac{p_{0}-p_{1}}{M}-x\right) \frac{%
d^{3}p}{p_{0}}  \nonumber \\
&=&\frac{1}{2}\cos \eta \int G(p_{0})\left( x-\frac{p_{T}^{2}/2}{M\left(
p_{0}+m\right) }\right) \delta \left( \frac{p_{0}+p_{1}}{M}-x\right) \frac{%
d^{3}p}{p_{0}}.  \nonumber
\end{eqnarray}%
So now we shall try to identify the transversity with the dimensionless
function%
\begin{equation}
\delta q(x)=\cos \eta \int G\left( p_{0}\right) \left( Mx-\frac{p_{T}^{2}/2}{%
p_{0}+m}\right) \delta \left( \frac{p_{0}+p_{1}}{M}-x\right) \frac{d^{3}p}{%
p_{0}}.  \label{n8}
\end{equation}%
If we use the alternative notation%
\begin{equation}
q(x)=f_{1}(x),\qquad \Delta q(x)=2g_{1}(x),\qquad \Delta q_{T}(x)=2g_{T}(x),
\label{m9}
\end{equation}%
then combination of the relations (\ref{tt8}) and (\ref{tt9a}) assuming $%
m\rightarrow 0$ gives%
\begin{equation}
\delta q(x)=\Delta q(x)+\Delta q_{T}(x).  \label{ma9}
\end{equation}%
Note that for the first moments, assuming the validity of the
Burkhardt-Cottingham sum rule, this implies 
\begin{equation}
\int_{0}^{1}\delta q(x)dx=2\int_{0}^{1}\Delta q(x)dx.  \label{2}
\end{equation}%
Further, using the Wanzura-Wilczek \cite{wawi} relation, which was proved
for our $g_{1},g_{2}$ in \cite{zav5}, 
\begin{equation}
g_{1}(x)+g_{2}(x)=\int_{x}^{1}\frac{g_{1}(y)}{y}dy,  \label{mb9}
\end{equation}%
the rule (\ref{ma9})\ can be represented as%
\begin{equation}
\delta q(x)=\Delta q(x)+\int_{x}^{1}\frac{\Delta q(y)}{y}dy.  \label{mc9}
\end{equation}%
Moreover, in the same paper we suggested the relations between the spin
functions and valence quark distributions:%
\begin{equation}
g_{1}^{q}(x)=\frac{\cos \eta _{q}}{2}\left[ q_{V}(x)-2x^{2}\int_{x}^{1}\frac{%
q_{V}(y)}{y^{3}}dy\right] ,\qquad g_{2}^{q}(x)=\frac{\cos \eta _{q}}{2}\left[
-\allowbreak \allowbreak q_{V}(x)+3x^{2}\int_{x}^{1}\frac{q_{V}(y)}{y^{3}}dy%
\right] ;\qquad q=u,d,  \label{t83}
\end{equation}%
which imply%
\begin{equation}
\delta q(x)=\cos \eta _{q}\left( q_{V}(x)-x^{2}\int_{x}^{1}\frac{q_{V}(y)}{%
y^{3}}dy\right) .  \label{t84}
\end{equation}%
So now one can calculate $\delta q$ either using experimental input on $%
\Delta q$ or from some fit on the valence distribution $q_{V}$. In Fig. \ref%
{fg2} we show transversities calculated with the use of formulae (\ref{mc9}%
) and (\ref{t84}). The spin functions $\Delta u$ and $\Delta d$ are here
extracted from the parameterization of the world data on the proton spin
function $g_{1}$\cite{e155g1} assuming the $SU(6)$ approach,\ 
\begin{equation}
\Delta u(x):\Delta d(x)=\frac{4}{3}:(-\frac{1}{3});\quad g_{1}(x)=\frac{1}{2}%
\left( \frac{4}{9}\Delta u(x)+\frac{1}{9}\Delta d(x)\right) .  \label{ta84}
\end{equation}%
For the valence functions $xu_{V}(x)$ and $xd_{V}(x)$ we use the
parametrization obtained by the standard global analysis in \cite{msr}. All
the parameterizations are taken for $Q^{2}=4GeV^{2}/c^{2}$. The  reason why
in this figure the (dashed) curve based on the experimental input on $g_{1}$
is above the (solid) curve calculated from fitted $q_{V}$, can be the same,
as it was discussed in \cite{zav5} directly for the $g_{1}$.

Now we check if the obtained transversities satisfy the Soffer inequality %
\cite{soffer}:%
\begin{equation}
\left| \delta q(x)\right| \leq \frac{1}{2}\left( q(x)+\Delta q(x)\right)
=q^{+}(x).  \label{tb84}
\end{equation}%
After inserting $\delta q,f_{1},g_{1}$ from the relations (\ref{n8}),(\ref%
{tt10}),(\ref{tt8}) we get%
\begin{eqnarray}
&&\left| \cos \eta _{q}\int G\left( p_{0}\right) \left( Mx-\frac{p_{T}^{2}/2%
}{p_{0}+m}\right) \delta \left( \frac{p_{0}+p_{1}}{M}-x\right) \frac{d^{3}p}{%
p_{0}}\right|  \label{n15} \\
&\leq &\frac{1}{2}\int \left( MxG(p_{0})+\cos \eta _{q}MxG(p_{0})-\cos \eta
_{q}\frac{p_{T}^{2}}{p_{0}+m}\right) \delta \left( \frac{p_{0}+p_{1}}{M}%
-x\right) \frac{d^{3}p}{p_{0}}  \nonumber
\end{eqnarray}%
and after rearranging the r.h.s. one obtains%
\begin{eqnarray}
&&\left| \cos \eta _{q}\int G\left( p_{0}\right) \left( Mx-\frac{p_{T}^{2}/2%
}{p_{0}+m}\right) \delta \left( \frac{p_{0}+p_{1}}{M}-x\right) \frac{d^{3}p}{%
p_{0}}\right|  \label{n16} \\
&\leq &\frac{1}{2}Mx\left( 1-\cos \eta _{q}\right) \int G(p_{0})\delta
\left( \frac{p_{0}+p_{1}}{M}-x\right) \frac{d^{3}p}{p_{0}}+\cos \eta
_{q}\int G\left( p_{0}\right) \left( Mx-\frac{p_{T}^{2}/2}{p_{0}+m}\right)
\delta \left( \frac{p_{0}+p_{1}}{M}-x\right) \frac{d^{3}p}{p_{0}},  \nonumber
\end{eqnarray}%
which means that%
\begin{equation}
\left| \delta q(x)\right| -\delta q(x)\leq Mx\sin ^{2}\left( \eta
_{q}/2\right) \int G(p_{0})\delta \left( \frac{p_{0}+p_{1}}{M}-x\right) 
\frac{d^{3}p}{p_{0}},  \label{n17}
\end{equation}%
which is apparently correct for $\delta q(x)\geq 0$. Further, since%
\begin{eqnarray*}
Mx-\frac{p_{T}^{2}/2}{p_{0}+m} &=&\frac{2Mxp_{0}+2Mxm-\left(
p_{0}^{2}-m^{2}\right) +p_{1}^{2}}{2\left( p_{0}+m\right) } \\
&=&\frac{2Mxp_{0}+2Mxm-\left( p_{0}^{2}-m^{2}\right)
+M^{2}x^{2}-2Mxp_{0}+p_{0}^{2}}{2\left( p_{0}+m\right) } \\
&=&\frac{\left( m+Mx\right) ^{2}}{2\left( p_{0}+m\right) }>0,
\end{eqnarray*}%
we have also%
\begin{eqnarray}
&&\int G\left( p_{0}\right) \left( Mx-\frac{p_{T}^{2}/2}{p_{0}+m}\right)
\delta \left( \frac{p_{0}+p_{1}}{M}-x\right) \frac{d^{3}p}{p_{0}}
\label{tt12} \\
&=&\int G\left( p_{0}\right) \frac{\left( m+Mx\right) ^{2}}{2\left(
p_{0}+m\right) }\delta \left( \frac{p_{0}+p_{1}}{M}-x\right) \frac{d^{3}p}{%
p_{0}}>0,  \nonumber
\end{eqnarray}%
which means that the transversity sign is controlled only by the sign of $%
\cos \eta $, which is determined by the sign of $G_{+}(p_{0})-G_{-}(p_{0})$.
So in our $SU(6)$ approach the inequality (\ref{n17}) is safely satisfied
for $u-$quarks. Now let us consider negative $\delta q$, for $d-$quarks in
the $SU(6)$ approach, when $\cos \eta =-1/3$ and $\sin ^{2}\left( \eta
/2\right) =2/3$. Then the combination of Eq. (\ref{n8}) and relation (\ref%
{n17}) gives%
\[
\int G\left( p_{0}\right) \left( Mx-\frac{p_{T}^{2}/2}{p_{0}+m}\right)
\delta \left( \frac{p_{0}+p_{1}}{M}-x\right) \frac{d^{3}p}{p_{0}}\leq Mx\int
G(p_{0})\delta \left( \frac{p_{0}+p_{1}}{M}-x\right) \frac{d^{3}p}{p_{0}}, 
\]%
which is obviously valid. So in the $SU(6)$ approach the Soffer inequality
is satisfied for both $u$ and $d$ quarks. However, let us now consider
rather extreme case when $\cos \eta \rightarrow -1$. Relation (\ref{n17})
for the angle $\eta =\pi $ reads%
\begin{equation}
2\int G\left( p_{0}\right) \left( Mx-\frac{p_{T}^{2}/2}{p_{0}+m}\right)
\delta \left( \frac{p_{0}+p_{1}}{M}-x\right) \frac{d^{3}p}{p_{0}}\leq 0,
\label{tt13}
\end{equation}%
which contradicts the inequality (\ref{tt12}), so in this limit also the
transversity (\ref{n8}) contradicts the Soffer inequality. Why?

The reason is in the assumption that transversity is generated by the same
function $H=G\cos \eta $ as the spin functions $g_{1}$ and $g_{2}$. The
resulting transversity contradicts the Soffer inequality in the case of
large negative quark polarization. Indeed, inequality (\ref{tb84}) means
that $|\delta q(x)|$ cannot exceed $q^{+}(x)$. At the same time, large
negative polarization takes place for $\cos \eta \rightarrow -1$; then $%
q^{+}(x)$ becomes small, while $\delta q(x)$ is large (and negative). Below
we shall modify the transversity definition as follows. The structure
functions are proportional (see e.g. \cite{GJJ}) to the combinations of
amplitudes:%
\begin{equation}
f_{1}\propto \sum_{X}\left( a_{++}^{\ast }(X)a_{++}(X)+a_{+-}^{\ast
}(X)a_{+-}(X)\right)  \label{tt15}
\end{equation}%
\begin{equation}
g_{1}\propto \sum_{X}\left( a_{++}^{\ast }(X)a_{++}(X)-a_{+-}^{\ast
}(X)a_{+-}(X)\right)  \label{tt16}
\end{equation}%
\begin{equation}
\delta q\propto \sum_{X}\left( a_{++}^{\ast }(X)a_{--}(X)+a_{--}^{\ast
}(X)a_{++}(X)\right) .  \label{tt17}
\end{equation}%
Now, in our approach we identify%
\begin{equation}
\sum_{X}a_{++}^{\ast }(X)a_{++}(X)=G_{+}(p_{0}),\qquad \sum_{X}a_{+-}^{\ast
}(X)a_{+-}(X)=G_{-}(p_{0}),  \label{tt18}
\end{equation}%
where $G_{+}\pm G_{-}$ are the distributions in relations (\ref{tt7}), from
which the structure functions $g_{1},g_{2},g_{T}$ and $f_{1}$ are
constructed in Eqs. (\ref{tt8})$-$(\ref{tt10}). But what about the remaining
interference function%
\begin{equation}
G_{T}(p_{0})\equiv \sum_{X}\left( a_{++}^{\ast }(X)a_{--}(X)+a_{--}^{\ast
}(X)a_{++}(X)\right) ,  \label{tta18}
\end{equation}%
which we are going to insert into Eq. (\ref{n8}) instead of $%
G_{+}-G_{-}=G\cos \eta $? The $G_{T}$\ is a new function, which has no
definite relation to the functions $G_{\pm }$. However, as a consequence of%
\begin{equation}
\sum_{X}\left\| a_{++}(X)\pm a_{--}(X)\right\| ^{2}\geq 0  \label{ttb18}
\end{equation}%
one gets%
\begin{equation}
\left| G_{T}(p_{0})\right| \leq G_{+}(p_{0})=G(p_{0})\cos ^{2}\left( \eta
/2\right) .  \label{ttc18}
\end{equation}%
So in the first step we check the Soffer inequality for both corresponding
extremes $\pm \delta q_{\max }(x);$ 
\begin{equation}
\delta q_{\max }(x)=\cos ^{2}\left( \eta _{q}/2\right) \int G(p_{0})\left(
Mx-\frac{p_{T}^{2}/2}{p_{0}+m}\right) \delta \left( \frac{p_{0}+p_{1}}{M}%
-x\right) \frac{d^{3}p}{p_{0}}.  \label{tt23}
\end{equation}%
After inserting into the Soffer inequality\footnote{%
One can start from relation (\ref{n16}), where on the l.h.s. $\cos \eta _{q}$
is substituted by $\cos ^{2}\left( \eta _{q}/2\right) $.} one gets for both
extremes 
\begin{equation}
0\leq \sin ^{2}\left( \eta _{q}/2\right) \int G(p_{0})\frac{p_{T}^{2}/2}{%
p_{0}+m}\delta \left( \frac{p_{0}+p_{1}}{M}-x\right) \frac{d^{3}p}{p_{0}},
\label{tt24}
\end{equation}%
so the inequality is satisfied for any transversity $\delta q(x)$ in the
band $\pm \delta q_{\max }(x)$ given by Eq. (\ref{tt23}) with {\it any} $%
\eta _{q}$. In fact, two inequalities are now satisfied: 
\begin{equation}
\left| \delta q(x)\right| \leq \delta q_{\max }(x)\leq \frac{1}{2}\left(
q(x)+\Delta q(x)\right)  \label{tz24}
\end{equation}%
and in this way we have shown that taking into account the interference
nature of the transversity at the stage of parton-hadron transition is quite
substantial for general compliance with the Soffer inequality.

Obviously, nothing can be said about saturation of the inequality%
\begin{equation}
\left| \delta q(x)\right| \leq \delta q_{\max }(x)  \label{tv24}
\end{equation}
within this simple approach. Concerning the sign of the transversity $\delta
q$, let us note, that now there is no simple correspondence with the sign of 
$\Delta q$.

Since the relations (\ref{n8}) and (\ref{tt23}) differ only in the $\eta -$
dependent factor ahead of the integral, the relations (\ref{ma9}),(\ref{mc9}%
) and (\ref{t84}) imply for the second approach%
\begin{equation}
\delta q_{\max }(x)=\kappa \cdot \left( \Delta q(x)+\Delta q_{T}(x)\right)
=\kappa \cdot \left( \Delta q(x)+\int_{x}^{1}\frac{\Delta q(y)}{y}dy\right)
;\qquad \kappa =\frac{\cos ^{2}\left( \eta _{q}/2\right) }{\cos \eta _{q}},
\label{tx24}
\end{equation}%
\begin{equation}
\delta q_{\max }(x)=\cos ^{2}\left( \eta _{q}/2\right) \left(
q_{V}(x)-x^{2}\int_{x}^{1}\frac{q_{V}(y)}{y^{3}}dy\right) .  \label{ty24}
\end{equation}

\begin{figure}[tbp]
\begin{center}
\epsfig{file=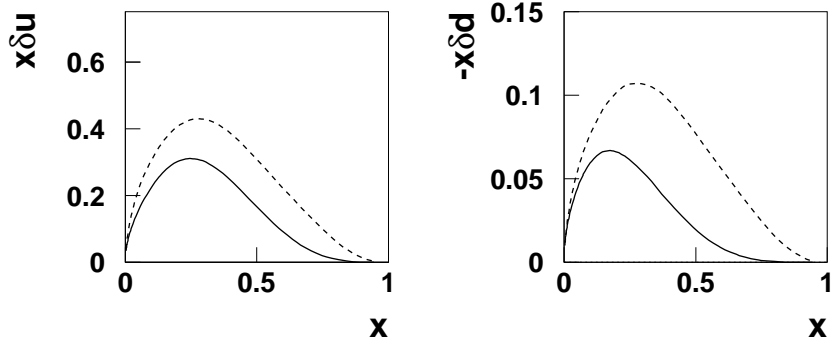, height=6cm}
\end{center}
\caption{Transversities of the $u$ and $d$ valence quarks calculated from
the valence distributions {\it (solid lines)} and extracted from the
experimental data on proton spin function $g_{1}$ {\it (dashed lines) }$-$
the first approach, see text.}
\label{fg2}
\end{figure}

\begin{figure}[tbp]
\begin{center}
\epsfig{file=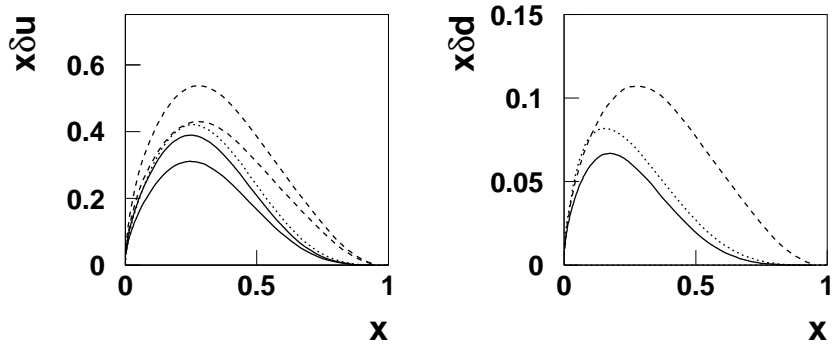, height=6cm}
\end{center}
\caption{Transversities of the $u$ valence quarks {\it (left)} calculated
from the valence distributions {\it (solid lines)} and extracted from the
experimental data on the proton spin function $g_{1}$ {\it (dashed lines)}.
Lower curves correspond to the first approach from Fig. \ref{fg2}, upper
curves represent the second approach given by $\protect\delta q_{\max }$
calculated from Eqs. (\ref{tx24}), (\ref{ty24}). The corresponding
transversities $\protect\delta q_{\max }$ of the second approach for the $d$
valence quarks {\it (right) }coincide, up to sign, with the first approach
from Fig. \ref{fg2}. The dotted lines represent the calculation %
\protect\cite{bochum}, with opposite sign for $d$ quarks.}
\label{fg3}
\end{figure}

This approach for the transversity is compared with the previous one in Fig. %
\ref{fg3}, again assuming $SU(6)$ approximation for contributions from $u$
and $d$ valence quarks. However, one should point out that curves
corresponding to the second approach represent only upper limits $\delta
q_{\max }$ for transversities, in the sense of the relation (\ref{tz24}).
The left part of the figure shows results for $d$-quarks. The relations (\ref%
{tx24}) {\it - dashed} and (\ref{ty24}) {\it - solid }curves are compared
with those in Fig. \ref{fg2}, calculated from Eqs. (\ref{mc9}),(\ref{t84}).
It follows, that curves of the second approach are enhanced by the factor $%
\cos ^{2}\left( \eta _{u}/2\right) /$ $\cos \eta _{u}=5/4$ with respect to
the first one. The right part of the figure demonstrates similar curves for $%
u-$quarks, but since $\cos ^{2}\left( \eta _{d}/2\right) =-$ $\cos \eta
_{d}=1/3,$ both the corresponding pairs of curves are equal up to sign. So
only the second pair is displayed. The dotted curves in both parts represent
tranversities calculated in Ref. \cite{bochum} in the LO and evolved from
the initial scale $0.6GeV^{2}/c^{2}$ to $4GeV^{2}/c^{2}$. Obviously, our
results are well compatible with them.

Further, let us analyze the relation%
\begin{equation}
\delta q_{\max }(x)\leq \frac{1}{2}\left( q(x)+\Delta q(x)\right) =q^{+}(x).
\label{tu24}
\end{equation}%
Obviously, its saturation is equivalent to the equality in relation (\ref%
{tt24}), which takes place either for $\eta =0$ (pure state of the quark
with polarization +) or for static quarks ($p_{T}^{2}=0$). On the other
hand, in the limit $m\rightarrow 0$ and with the use of Eq. (\ref{t83}) one
can write 
\begin{equation}
q^{+}(x)=\frac{1}{2}\left[ q_{V}(x)+\cos \eta _{q}\left(
q_{V}(x)-2x^{2}\int_{x}^{1}\frac{q_{V}(y)}{y^{3}}dy\right) \right] =\cos
^{2}\left( \eta _{q}/2\right) \cdot q_{V}(x)-x^{2}\cos \eta _{q}\int_{x}^{1}%
\frac{q_{V}(y)}{y^{3}}dy.  \label{t25}
\end{equation}

In this way, both sides of relation (\ref{tu24}) are displayed in Fig. \ref%
{fg4}. It is seen that particularly the transversity of $d$-quarks in this
approach is constrained considerably more than only by the Soffer bound $%
q^{+}$. The relationship between both bounds depends on $\cos \eta _{q}$.
One can check that%
\begin{equation}
\frac{\int_{0}^{1}\delta q_{\max }(x)dx}{\int_{0}^{1}q^{+}(x)dx}=\frac{%
2+2\cos \eta _{q}}{3+\cos \eta _{q}},  \label{t26}
\end{equation}%
which in the $SU(6)$\ approach gives fractions $10/11$ and $1/2$ for $u$ and 
$d$ quarks.

\section{Summary}

The covariant QPM, which takes into account intrinsic quark motion, was
generalized to involve the transversity distribution. Two ways of
generalization were considered:

1) The interference effects were assumed on a quark level only (interference
of vector and scalar currents produce the quark trace with one Lorentz
index) and the generic quark polarized distribution $H=G_{+}-G_{-}$ was
assumed the same as that for the structure functions $g_{1}$\ and $g_{2}$.
We derived the relation between the transversity $\delta q$ and the usual
polarized distribution $\Delta q$, which implies that the resulting
transversity is roughly twice as large as the usual distribution function.
We discussed the compatibility of the obtained transversity with the Soffer
inequality and we have found that the inequality is violated in the case of
large negative quark polarization.

2) In the second approach we accounted also for interference effects at the
quark-hadron transition stage, which is formally represented by the
interference sum (\ref{tta18}). In this approach, assuming the validity of
Soffer inequality at the stage of parton-hadron transition (\ref{ttc18}), we
obtained a new bound on the transversity, rather than the transversity in
itself. This bound is more strict, than the Soffer one. Roughly speaking,
the new bound is more restrictive for quarks with a greater proportion of
intrinsic motion and/or smaller (or negative) portion in the resulting
polarization.

For both approaches, with the use of the obtained sum rules, we have done
the numeric calculations, in which either experimentally measured spin
function $g_{1}(x)$ or the valence quark distributions $q_{V}(x)$ were used
as an input. To conclude, in general our calculations suggest that the quark
intrinsic motion plays an important role for the spin functions - including
the transversity.

\newpage

\begin{figure}[tbp]
\begin{center}
\epsfig{file=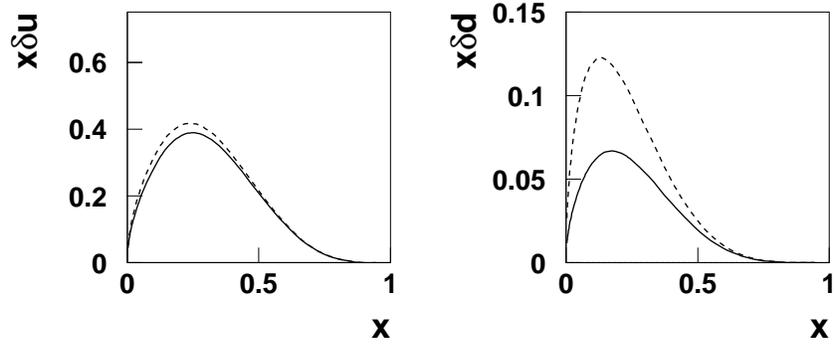, height=6cm}
\end{center}
\caption{Bounds on the transversities of the $u$ and $d$ valence quarks: $%
\protect\delta q_{\max }$ {\it (solid lines)} and $q^{+}$ {\it (dashed lines).%
} }
\label{fg4}
\end{figure}

\end{document}